\documentclass[a4paper, prl, twocolumn,superscriptaddress,showpacs]{revtex4}

\pdfoutput=1
\usepackage{amssymb}
\usepackage{amsmath}
\usepackage{epsfig}
\usepackage{color}
\usepackage{graphics, graphicx}
\usepackage{bbold}
\usepackage{psfrag}
\usepackage{mathcomp}
\usepackage{subfigure}
\usepackage{verbatim}
\usepackage{color}
\usepackage[colorlinks,citecolor=blue]{hyperref}

\begin{document}

\date{\today}
\title{Unconventional superfluid in a two-dimensional Fermi gas with anisotropic spin-orbit coupling and Zeeman fields}

\author{Fan Wu}
\affiliation{Key Laboratory of Quantum Information, University of Science and Technology of China,
CAS, Hefei, Anhui, 230026, People's Republic of China}
\author{Guang-Can Guo}
\affiliation{Key Laboratory of Quantum Information, University of Science and Technology of China,
CAS, Hefei, Anhui, 230026, People's Republic of China}
\author{Wei Zhang}
\email{wzhangl@ruc.edu.cn}
\affiliation{Department of Physics, Renmin University of China, Beijing 100872, People's Republic of China}
\author{Wei Yi}
\email{wyiz@ustc.edu.cn}
\affiliation{Key Laboratory of Quantum Information, University of Science and Technology of China,
CAS, Hefei, Anhui, 230026, People's Republic of China}
\begin{abstract}
We study the phase diagram of a two-dimensional ultracold Fermi gas with the synthetic spin-orbit coupling (SOC) that has recently been realized at NIST. Due to the coexistence of anisotropic SOC and effective Zeeman fields in the NIST scheme, the system shows rich structure of phase separation involving exotic gapless superfluid and Fulde-Ferrell-Larkin-Ovchinnikov (FFLO) pairing states with different center-of-mass momentum. In particular, we characterize the stability region of FFLO states and demonstrate their unique features under SOC. We then show that the effective transverse Zeeman field in the NIST scheme can qualitatively change the landscape of the thermodynamic potential which leads to intriguing effects such as the disappearance of pairing instability, the competition between different FFLO states, and the stabilization of a fully gapped FFLO state. These interesting features may be probed for example by measuring the {\it in-situ} density profiles or by the momentum-resolved radio-frequency spectroscopy.
\end{abstract}
\maketitle


\emph{Introduction}.--
The recent realization of synthetic spin-orbit coupling (SOC) in ultracold atoms has greatly extended the horizon of quantum simulation in these systems  \cite{gauge2exp,fermisocexp,shuaiexp}. Interesting phenomena like topological insulators, quantum spin Hall effects, topological superfluidity, etc., where SOC plays a key role \cite{kanereview}, may now be studied in ultracold Fermi gases. A great amount of efforts have recently been dedicated to the clarification of novel phases and phase transitions in a spin-orbit coupled, strongly interacting Fermi gas of ultracold atoms, where the interplay of SOC, pairing superfluidity and effective Zeeman fields can lead to exotic superfluid phases in various dimensions \cite{zhang,sato,soc3,chuanwei,soc4,soc6,iskin,thermo,2d2,2d1,melo,helianyi,wy2d,xiaosen,wypolaron,chuanweifflo,carlosnistsoc,iskinnistsoc}.

While most of these studies assume the Rashba SOC, only an equal Rashba and Dresselhaus (ERD) SOC has been realized in experiments \cite{gauge2exp,fermisocexp,shuaiexp,xiongjun}. With spatial anisotropy and effectively reduced SOC dimension, the NIST SOC can lead to qualitatively different superfluid phases compared to those under Rashba SOC \cite{carlosnistsoc,iskinnistsoc,shenoy}. The picture is further complicated by the existence of effective Zeeman fields, both axial and transverse, in the NIST scheme, which are tunable by adjusting the laser parameters \cite{gauge2exp}. On the other hand, although there has been discussions of Fulde-Ferrell-Larkin-Ovchinnikov (FFLO) state in these systems \cite{wypolaron,chuanweifflo,shenoy,puhantwobody}, the FFLO state has not been systemtaically characterized in the context of the NIST scheme.

In this work, we study the exotic superfluid phases in a two-dimensional (2D) ultracold Fermi gas near a wide Feshbach resonance and with the NIST SOC. Given the enhanced stability of FFLO states in two dimensions \cite{simon2dfflo}, as well as the recent development of {\it in-situ} detection schemes \cite{ccinsitu}, 2D Fermi gas is an ideal platform for the investigation of the unconventional pairing states, FFLO in particular, under SOC. By mapping out the zero-temperature phase diagram, we find that, in the absence of an effective transverse Zeeman field, typically two different gapless superfluid states can appear, similar to the case of a three-dimensional (3D) polarized Fermi gas with Rashba SOC \cite{iskin,thermo}. Furthermore, an FFLO state with center-of-mass momentum perpendicular to the direction of the anisotropic SOC can be stabilized. The quasi-particle (hole) dispersion spectra of this FFLO state feature non-trivial gapless contours in momentum space, distinct from that of the FFLO state without SOC. The unique dispersion spectra and the gapless contours may be detected via momentum-resolved radio-frequency (rf) spectroscopy \cite{dsjinarp}. We then discuss the influence of effective transverse Zeeman fields on the system. In general, a transverse Zeeman field eradicates the pairing instability, typical of an attractively interacting Fermi gas under SOC in the large polarization limit \cite{wy2d,sauNJP}, and stabilizes FFLO states with center-of-mass momentum along the direction of the anisotropic SOC. These lead to the appearance of normal state on the phase diagram, and the competition between various FFLO states with different center-of-mass momentum. As a result, first-order phase transitions are abundant on the phase diagram, which should leave signatures in the {\it in-situ} density profiles of a trapped gas. Finally, we identify a continuous transition between the gapless FFLO states and an interesting fully gapped FFLO state, which is the result of SOC-induced
spin mixing and Fermi surface deformation. With recent progress in the experimental investigation of 2D Fermi gases \cite{2dgasexpall} and the realization of SOC in a degenerate Fermi gas \cite{fermisocexp}, we expect that many features reported in this work will be observed in future experiments.

\emph{Model}.--
We consider a two-dimensional two-component Fermi gas with the NIST SOC \cite{gauge2exp,fermisocexp}, where the Hamiltonian can be written as
\begin{eqnarray}
H&=&\sum_{\mathbf{k},\sigma=\uparrow,\downarrow}\xi_{k}a^{\dag}_{\mathbf{k}\sigma}a_{\mathbf{k}\sigma}+\sum_{\mathbf{k}}h (a^{\dag}_{\mathbf{k}\uparrow}a_{\mathbf{k}\downarrow}+h.c.)\nonumber\\
&+&\sum_{\mathbf{k}}[(\alpha k_x -h_x) a^{\dag}_{\mathbf{k}\uparrow}a_{\mathbf{k}\uparrow}-(\alpha k_x-h_x) a^{\dag}_{\mathbf{k}\downarrow}a_{\mathbf{k}\downarrow}]\nonumber\\
&+&\frac{U}{{\cal S}}\sum_{\mathbf{k},\mathbf{k}',\mathbf{q}}a^{\dag}_{\mathbf{k}+\mathbf{q}\uparrow} a^{\dag}_{-\mathbf{k}+\mathbf{q}\downarrow}a_{-\mathbf{k}'+\mathbf{q}\downarrow}a_{\mathbf{k}'+\mathbf{q}\uparrow}.\label{eqnHorg}
\end{eqnarray}
Here, $\xi_{k}=\epsilon_k-\mu$, $\epsilon_k=\hbar^2k^2/2m$, $a_{\mathbf{k}\sigma}$ ($a^{\dag}_{\mathbf{k}\sigma}$) is the annihilation (creation) operator for the hyperfine spin state $\sigma$ with $\sigma=(\uparrow,\downarrow)$, $m$ is the atomic mass, ${\cal S}$ is the quantization volume in two dimensions, and $\alpha$ denotes the strength of the SOC. The effective Zeeman fields $h$ and $h_x$ are proportional to the effective Rabi-frequency and the two-photon detuning, respectively, of the Raman process in the experiment \cite{gauge2exp}. Note that the SOC terms in Hamiltonian (\ref{eqnHorg}) differ from the standard ERD form by a spin-rotation, and that we adopt the convention in the ERD form and label $h$ ($h_x$) as the axial (transverse) Zeeman field. The bare $s$-wave interaction rate $U$ should be renormalized as \cite{renorm}: $1/U=-1/{\cal S}\sum_{\mathbf{k}}1/(E_b+2\epsilon_{\mathbf{k}})$, where $E_b$ is the binding energy of the two-body bound state in two dimensions without SOC, which can be tuned, for instance, via the Feshbach resonance technique.

We focus on the zero-temperature properties of the Fulde-Ferrell (FF) pairing states with a single valued center-of-mass momentum on the mean-field level \cite{fflo}, which should provide a qualitatively correct phase diagram at zero temperature. The effective mean field Hamiltonian can then be arranged into a matrix form in the hyperfine spin basis $\left\{a_{\mathbf{k}\uparrow},a^{\dag}_{\mathbf{Q}-\mathbf{k}\uparrow},a_{\mathbf{k}\downarrow},a^{\dag}_{\mathbf{Q}-\mathbf{k}\downarrow}\right\}^{T}$ \begin{eqnarray}
H_{\text{eff}}&=&\frac{1}{2}\sum_{\mathbf{k}}\begin{pmatrix}
\lambda^{+}_k&0&h&\Delta_Q\\
0&-\lambda^{+}_{Q-k}&-\Delta^{\ast}_Q&-h\\
h&-\Delta_Q&\lambda^{-}_k&0\\
\Delta^{\ast}_Q&-h&0&-\lambda^{-}_{Q-k}
\end{pmatrix}\nonumber\\
&+&\sum_{\mathbf{k}}\xi_{|\mathbf{Q}-\mathbf{k}|}-\frac{|\Delta_Q|^2}{U},\label{eqnHeff}
\end{eqnarray}
where $\lambda_{k}^{\pm}=\xi_{k}\pm\alpha k_x\mp h_x$, the order parameter $\Delta_Q=U\sum_{\mathbf{k}}\left\langle a_{\mathbf{Q}-\mathbf{k}\downarrow} a_{\mathbf{k}\uparrow} \right\rangle$. It is then straightforward to diagonalize the effective Hamiltonian and evaluate the thermodynamic potential at zero temperature
\begin{equation}
\Omega=\sum_{\mathbf{k}}\xi_{|\mathbf{Q}-\mathbf{k}|}+\sum_{\mathbf{k},\nu}\theta(-E^{\eta}_{\mathbf{k},\nu})E^{\eta}_{\mathbf{k},\nu}-\frac{|\Delta_Q|^2}{U},
\label{eqnOmega}
\end{equation}
where the quasi-particle (hole) dispersion $E^{\eta}_{\mathbf{k},\nu}$ ($\nu=1,2$, $\eta=\pm$) are the eigenvalues of the matrix in Hamiltonian (\ref{eqnHeff}), and $\theta(x)$ is the Heaviside step function. Without loss of generality, we assume $h,h_x>0$, $\Delta_0=\Delta$, and $\Delta_Q$ to be real throughout the work. The pairing order parameter $\Delta_{\mathbf{Q}}$ as well as the center-of-mass momentum $\mathbf{Q}$ for the pairs can then be found by minimizing the thermodynamic potential in Eq. (\ref{eqnOmega}).

In general, the Hamiltonian (\ref{eqnHeff}) cannot be diagonalized analytically, and the thermodynamic potential needs to be evaluated numerically. However, for pairing states with zero center-of-mass momentum ($Q=0$), analytical form of the dispersion relation can be obtained along the $k_x=0$ axis, with: $E^{\pm}_{k_y,1}=\left|\sqrt{\xi_{k_y}^2+\Delta^2}\pm\sqrt{h^2+h_x^2}\right|$, and $E^{\pm}_{k_y,1}=-E^{\pm}_{k_y,2}$. Apparently, the branches $E^{-}_{k_{y},\nu}$ can cross zero for finite $\Delta$, which leads to gapless superfluid phases with isolated gapless points in momentum space, similar to the case of a 3D Fermi gas with Rashba SOC. We find numerically that for finite $\Delta$, the gapless points can only occur on the $k_x=0$ axis \cite{footnoteongaplesspoint}, therefore it is sufficient to study the phase boundaries based on the dispersion along this axis. Typically, there can be two or four gapless points on the axis, which are symmetric with respect to the origin. As the location of the gapless points ($k_{yc}$) must satisfy the relation $E^{-}_{k_{yc},\nu}=0$, it is easy to see that there are four gapless points when $\mu>\sqrt{h^2+h_x^2-\Delta^2}$ and $h^2+h_x^2>\Delta^2$, two gapless points when $|\mu|<\sqrt{h^2+h_x^2-\Delta^2}$ and $h^2+h_x^2>\Delta^2$. Similar to the case of a 3D Fermi gas with Rashba SOC \cite{thermo}, we may then identify nodal superfluid states with two gapless points (nSF1) and those with four gapless points (nSF2) as distinct gapless superfluid phases. As has been pointed out in Ref. \cite{carlosnistsoc}, these phases have locally non-trivial topological properties in momentum space near a given gapless point.

\begin{figure}[tbp]
\includegraphics[width=8.6cm]{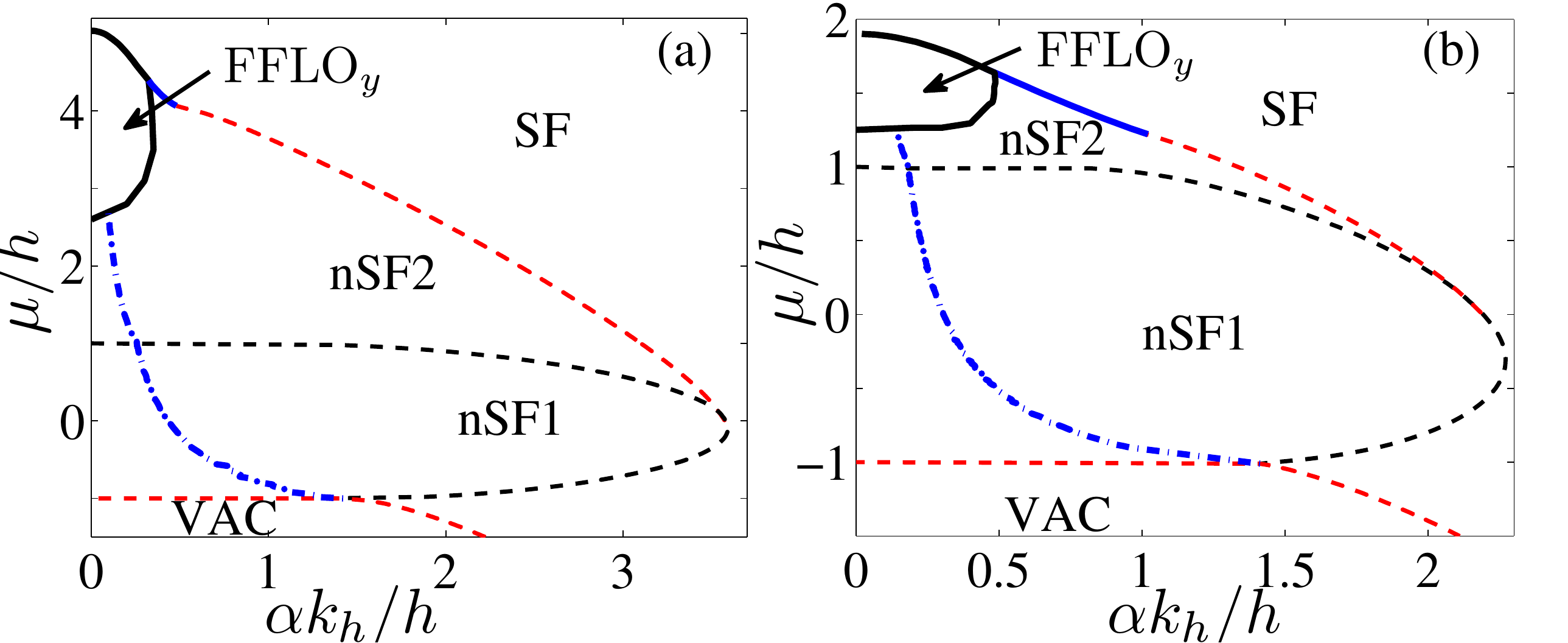}
\caption{(Color online) Phase diagrams on the $\mu$--$\alpha$ plane for (a) $E_b/h=0.2$, $h_x=0$, and (b) $E_b/h=0.5$, $h_x=0$. The solid curves are first-order boundaries, while the dashed curves represent phase boundaries of continuous phase transitions. The dash-dotted curves traversing the nSF phases are the threshold with $\Delta/h=10^{-3}$. The axial Zeeman field $h$ is taken to be the unit of energy, while the unit of momentum $k_h$ is defined through $\hbar^2k_h^2/2m=h$. }
\label{fig1}
\end{figure}

\emph{Without effective transverse Zeeman fields}.--
We first examine the phases in the absence of effective transverse Zeeman fields. With $h_x=0$, the Hamiltonian (\ref{eqnHeff}) can be diagonalized analytically for pairing states with $Q=0$. The dispersion spectra are: $E^{\pm}_{\mathbf{k},1}=\sqrt{\xi^2_{k}+\Delta^2+h^2+\alpha^2k_x^2\pm2E_0}$, and $E^{\pm}_{\mathbf{k},1}=-E^{\pm}_{\mathbf{k},2}$, where $E_0=\sqrt{\xi_k^2(h^2+\alpha^2k_x^2)+\Delta^2h^2}$. While for pairing states with finite center-of-mass momentum, one must resort to numerical diagonalization. With these, we map out the phase diagram on the $\mu$--$\alpha$ plane with fixed $h$ (see Fig. \ref{fig1}). Under the local density approximation, the phases traversed by a downward vertical line in the diagram represent those one should encounter starting from a trap center and moving to its edge.

From Fig. \ref{fig1}, we see that the topological superfluid phase in a 2D polarized Fermi gas with Rashba SOC is now replaced by nSF1 and nSF2, as we have anticipated. On the other hand, similar to the case of a 2D Fermi gas with Rashba SOC, pairing state exists even in the large polarization limit \cite{wy2d,wypolaron}. This indicates the persistence of pairing instability despite the anisotropy of the NIST SOC. Similar to the Rashba case, the pairing instability is dictated by the existence of singularities in the gap equation when $\Delta=0$ \cite{sauNJP,wy2d}, which is equivalent to the condition that the dispersion spectra should cross zero at $\Delta=0$, i.e., $\xi_k^2=\alpha^2k_x^2+h^2$. As this is also the condition for the existence of Fermi surfaces in a normal state (N) under SOC, the normal state is absent from the phase diagram and vacuum (VAC) shows up in
its place.

\begin{figure}[tbp]
\includegraphics[width=4.2cm]{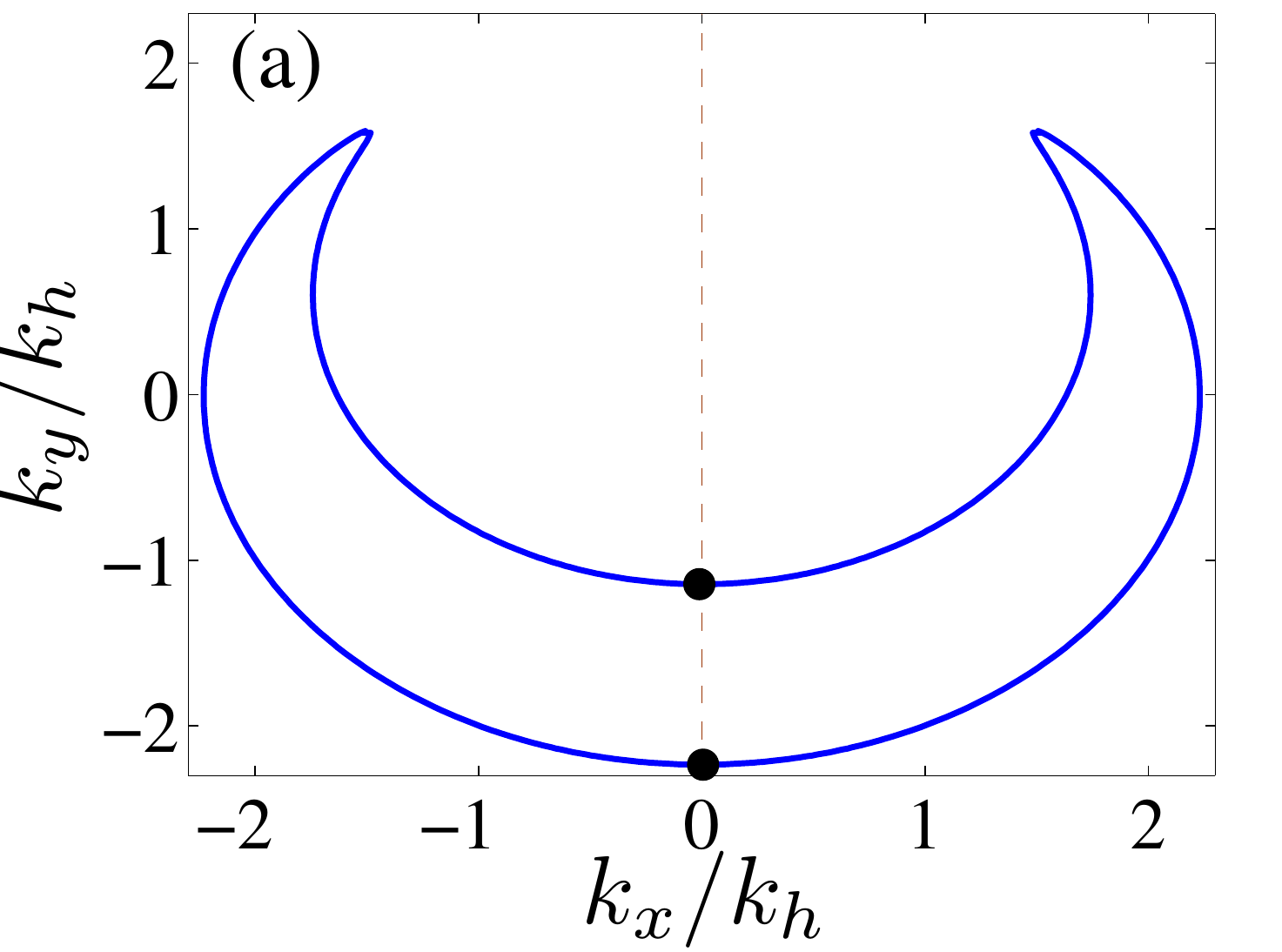}
\includegraphics[width=4.2cm]{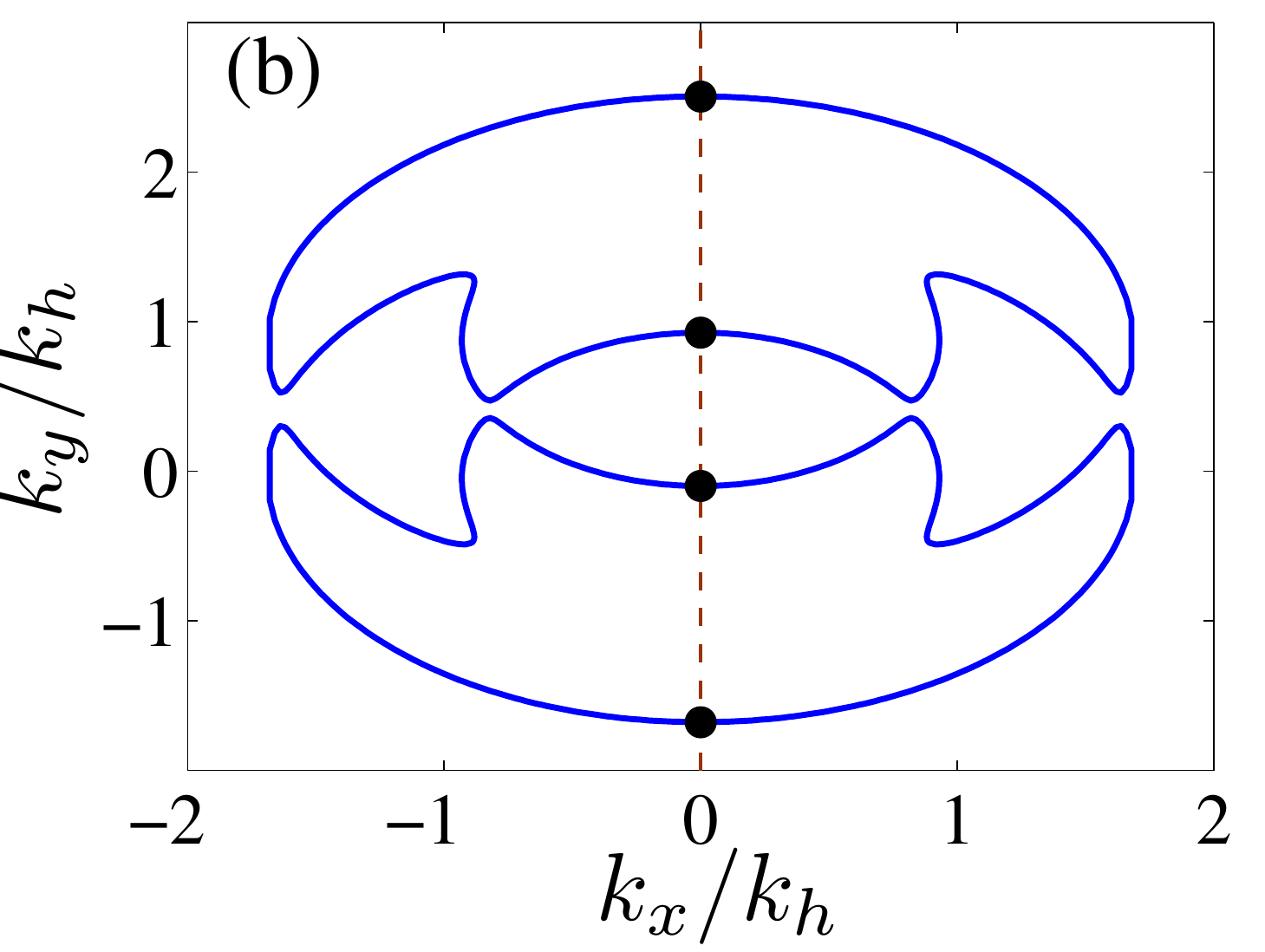}
\includegraphics[width=4.2cm]{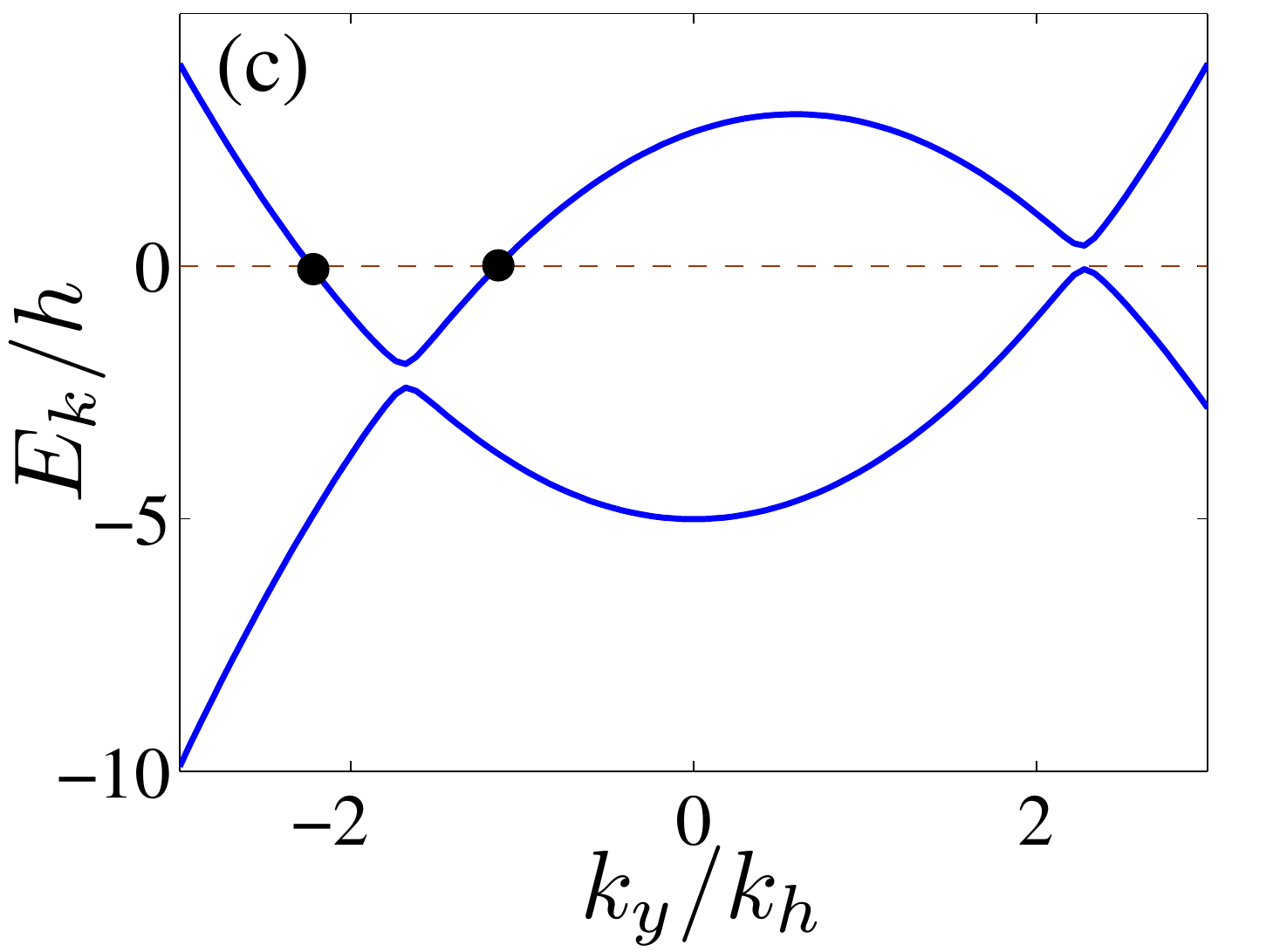}
\includegraphics[width=4.2cm]{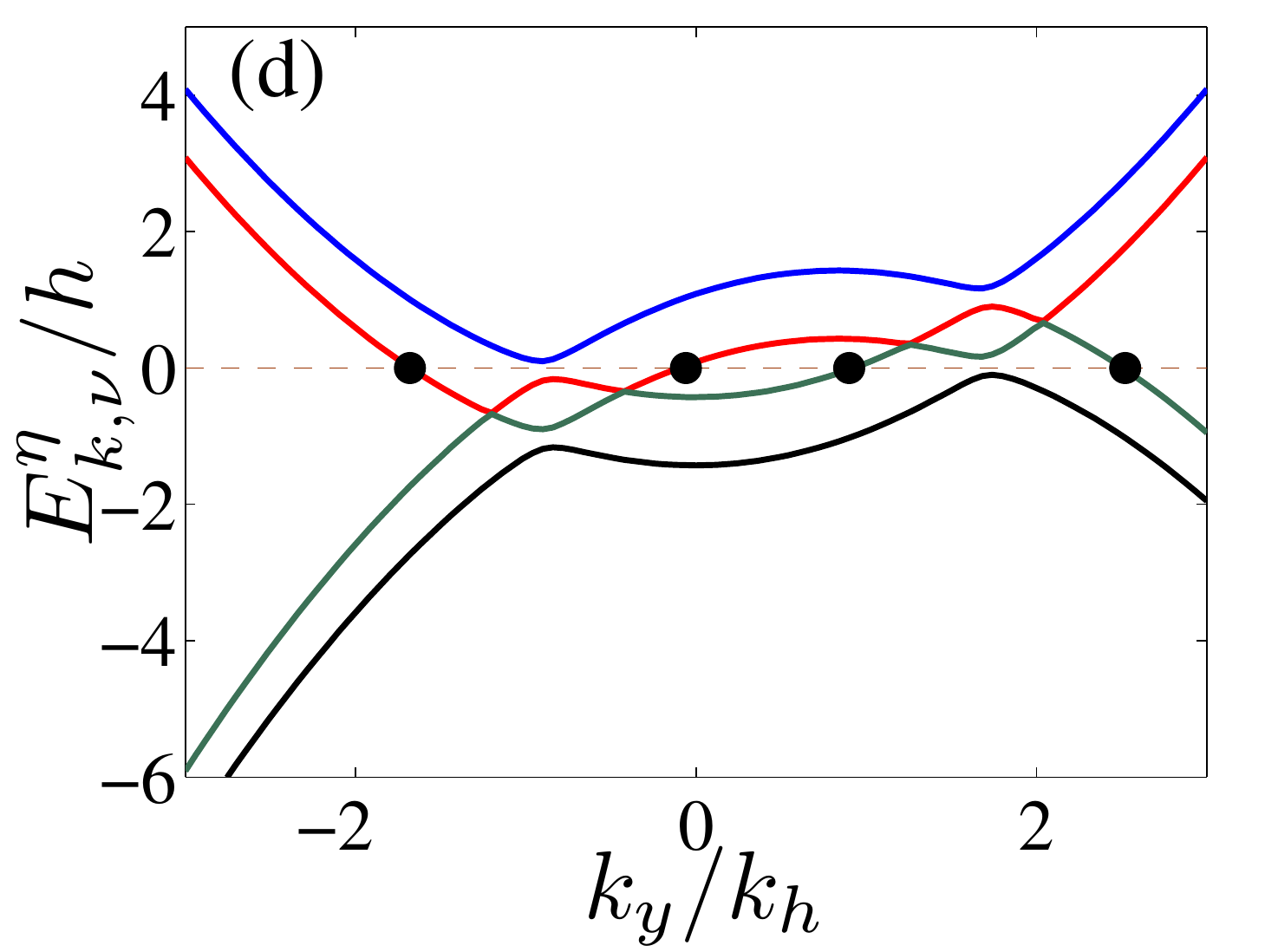}
\caption{(Color online) (a) Contour of gapless points in momentum space for a standard FFLO state in a polarized Fermi gas without SOC. (b) Contours of gapless points in momentum space for a typical FFLO$_y$ state with NIST SOC and with $h_x=0$. (c) Dispersion spectra along the $k_x=0$ axis for the FFLO state in (a). (d) Dispersion spectra along the $k_x=0$ axis for the FFLO$_y$ state in (b). The parameters for (b)(d) are: $E_b/h=0.5$, $\alpha k_h/h=0.21$, $\mu/h=1.83$, $\Delta_Q/h\sim0.28$, $Q_y/k_h\sim0.83$, $h_x=0$. Gapless points on the $k_x=0$ axis are indicated by bold dots.}
\label{fig2}
\end{figure}

\begin{figure}[tbp]
\includegraphics[width=6.5cm]{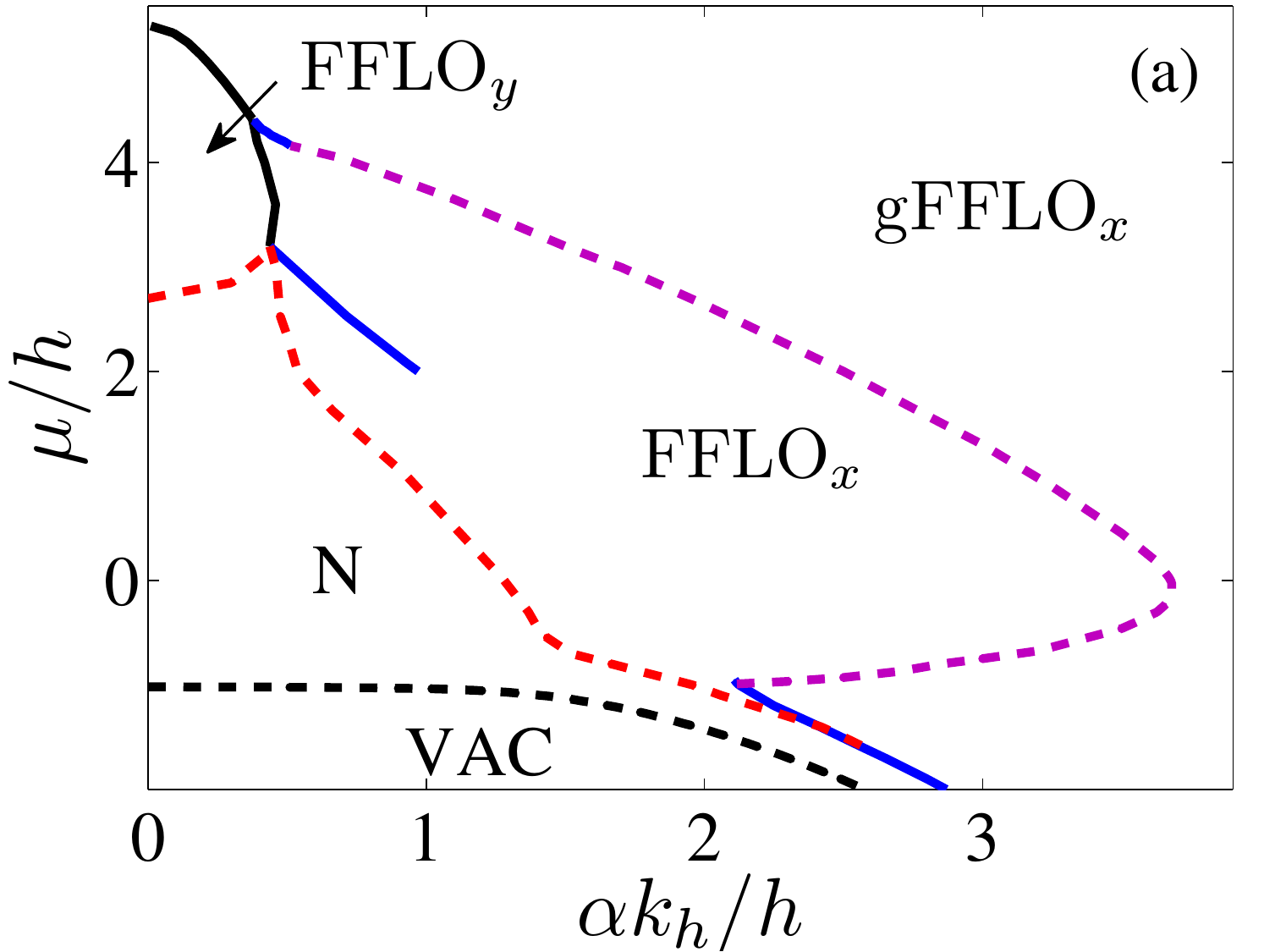}
\includegraphics[width=6.3cm]{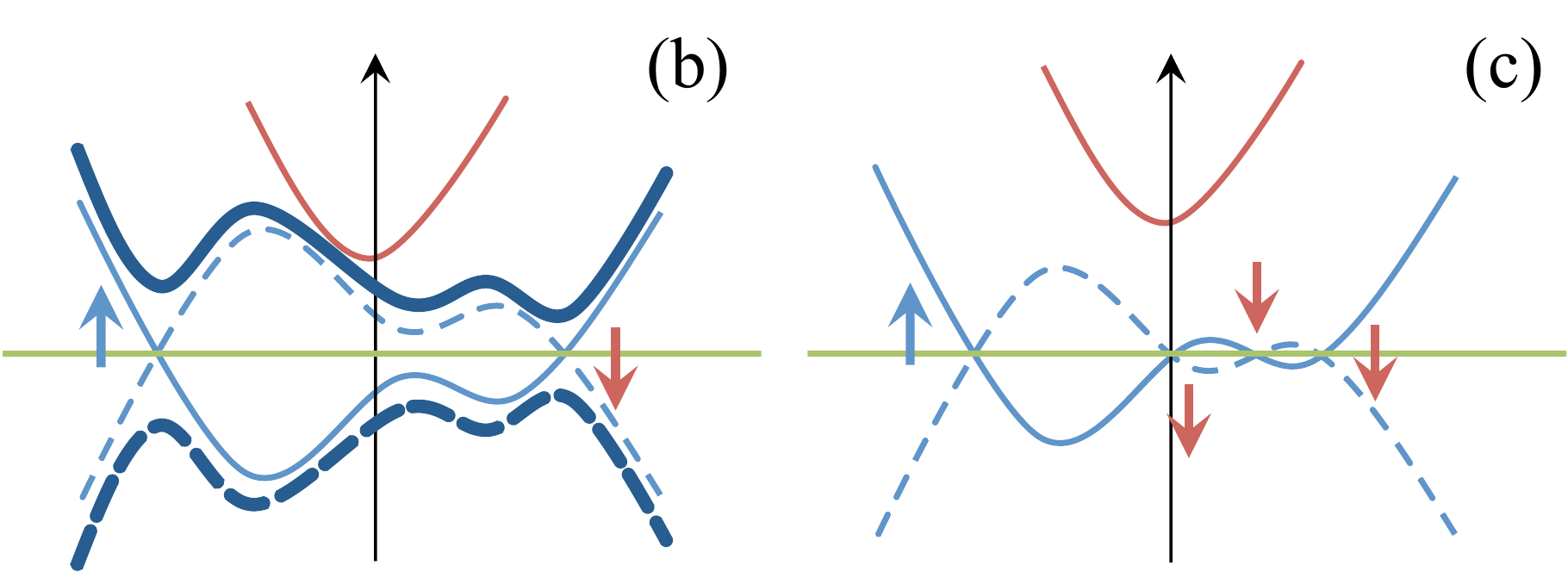}
\caption{(Color online) (a) Phase diagram on the $\mu$--$\alpha$ plane for $E_b/h=0.2$, $h_x/h=0.2$. The solid curves are first-order boundaries, the dashed curves represent continuous phase boundaries. (b)(c) Illustration of dispersion spectra and possible pairing states in the presence of $h_x$. While an FFLO$_x$ pairing state with a small and negative $Q_x$ is possible in (b), various FFLO$_x$ pairing states with different $Q_x$ are possible in (c). The thin (solid and dashed) curves represent the non-interacting bands formed by hyperfine states under SOC and an axial Zeeman field. The bold solid (dashed) curve represents the quasi-particle (hole) dispersion. The horizontal line represents the Fermi surface.  }\label{fig3}
\end{figure}

For pairing states with finite center-of-mass momentum $\mathbf{Q}$, we find that with $h_x=0$, $\mathbf{Q}$ lies along the $y$-axis (FFLO$_y$), i.e. perpendicular to the direction of the NIST SOC. Interestingly, the pairing states with $\mathbf{Q}$ along the $x$-axis (FFLO$_x$) are also stable against the normal state, but are metastable compared to the FFLO$_y$ state. The FFLO$_y$ state is bounded by the SF state from above and the nSF2 state from below, both via first-order phase boundaries. While the first-order boundary between the FFLO$_y$ state and the SF state is consistent with the mean field picture in the absence of SOC, the lower boundary against the nSF2 state is a direct result of pairing instability under SOC, as otherwise the nSF2 state at the boundary is replaced by the normal state and the phase boundary would become continuous. Indeed, the first-order boundary between FFLO$_y$ and nSF2 approaches a critical point at $\alpha=0$, which is a second-order critical point that can be solved analytically for a 2D polarized Fermi gas \cite{simon2dfflo}. As the SOC strength increases, the stability region of the FFLO$_y$ state decreases on the phase diagram, implying competition between FFLO$_y$ pairing and SOC.

To further characterize the properties of the FFLO$_y$ state, we demonstrate in Fig. \ref{fig2} typical gapless points and dispersion spectra of the FFLO$_y$ state in momentum space. For a standard FFLO state in a polarized 2D Fermi gas, the gapless points form a closed contour on the $k_x$--$k_y$ plane (see Fig. \ref{fig2}(a)). As SOC breaks the inversion symmetry, the dispersion spectra split into four branches with the symmetry $E^{\eta}_{k_x,k_y,\nu}=-E^{\eta'}_{k_x,Q_y-k_y,\nu'}$ (see Fig. \ref{fig2}(b)(d)). Hence two distinct contours of gapless points show up in momentum space. This is a unique feature of the FFLO$_y$ state in a Fermi gas with SOC, which may be probed using momentum resolved rf spectroscopy.

\emph{With an effective transverse Zeeman field}.--
We now turn to the case where an effective transverse Zeeman field $h_x$ is present. As illustrated in Fig. \ref{fig3}(b)(c), the transverse field makes the bands asymmetric. The Fermi surface becomes deformed, which may render $s$-wave pairing with zero center-of-mass momentum unfavorable even under SOC, i.e., pairing instability may no longer exist. This is reflected in the thermodynamic potential, where a local minimum can appear at $\Delta=0$. A more thorough examination shows an intriguing landscape, where the thermodynamic potential may have up to three local minima with $Q=0$, corresponding to the SF, the nSF2 and the normal state, respectively \cite{seesupp}. Hence, there can potentially be three different first-order transitions with $Q=0$.

The picture is further complicated when we take FF pairing states into account. With deformed Fermi surface, pairing states with center-of-mass momentum along the $x$-direction can be stabilized against the SF state. A qualitative understanding is that due to the Fermi surface asymmetry, the local minima in the case of $Q=0$ can be shifted to finite $Q_x$ in the presence of a finite $h_x$. The magnitude and direction of this shift depends on the Fermi surface asymmetry (see Fig. \ref{fig3}(b)(c)). Under certain conditions, there may even exist several different FFLO$_x$ states with distinct $Q_x$ and $\Delta_Q$ (see Fig. \ref{fig3}(c)). The ground state of the system therefore is the result of competitions between the various FFLO pairing states and the normal state.

\begin{figure}[tbp]
\includegraphics[width=8.6cm]{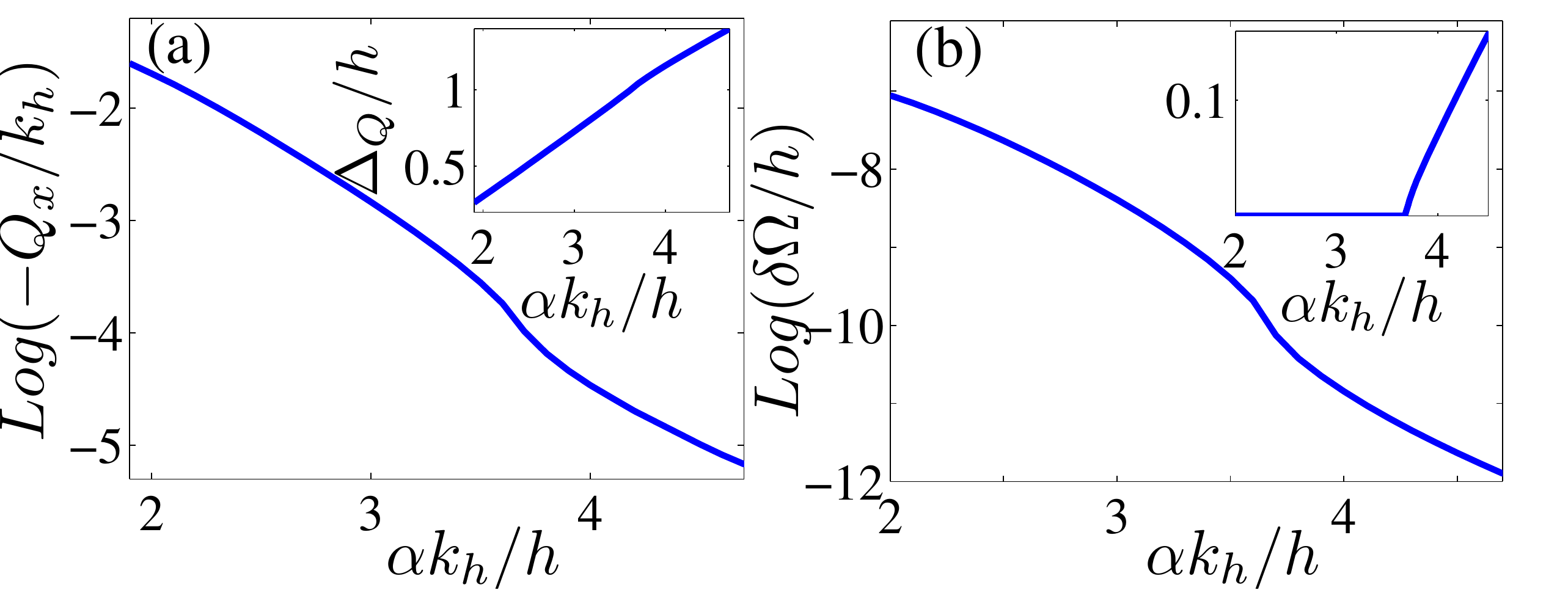}
\includegraphics[width=8.6cm]{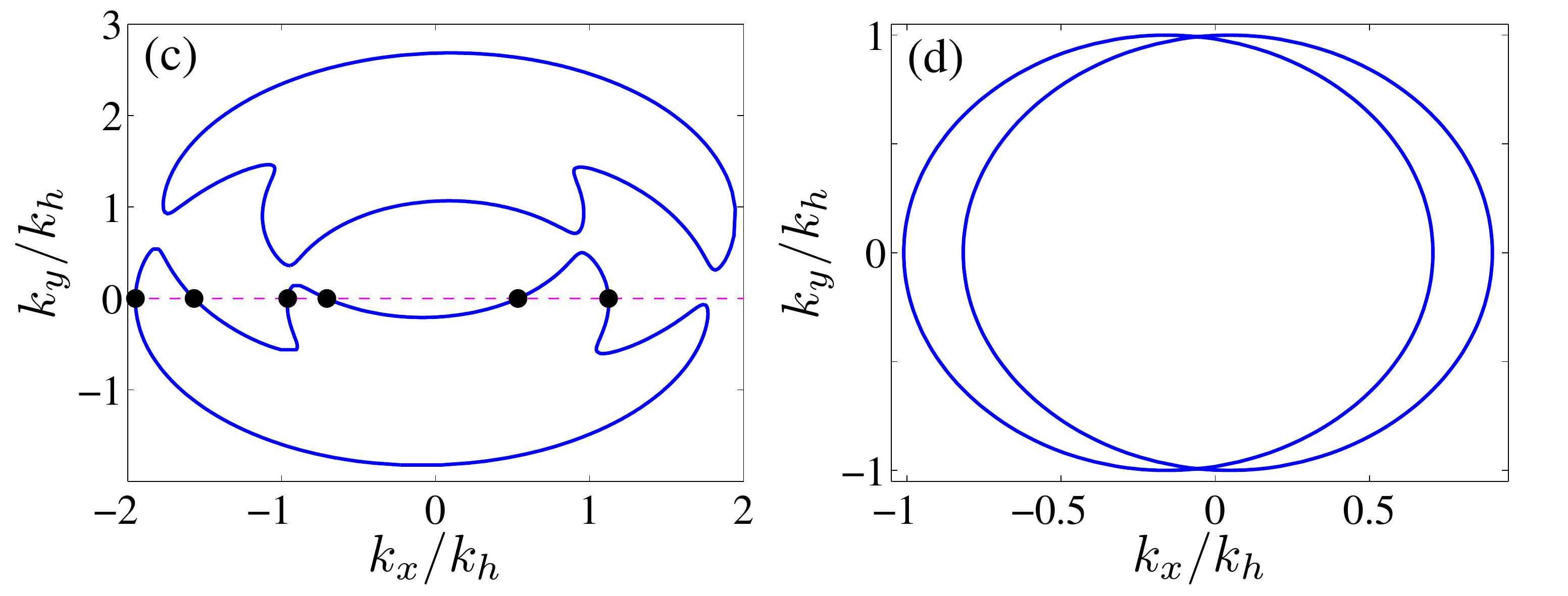}
\caption{(Color online) (a) Evolution of the center-of-mass momentum and pairing order parameter (inset) for the ground state as the SOC strength increases. (b) Evolution of the difference in thermodynamic potential between the SF state and the FFLO$_x$ state with increasing SOC strength, with $\delta\Omega=\Omega_{\rm SF}-\Omega_{\rm FFLO_x}$. The inset shows the evolution of the minimum excitation gap with increasing SOC strength. The parameters for (a)(b) are: $E_b/h=0.2$, $h_x/h=0.2$, $\mu=0$. (c) Typical gapless contours in momentum space for an FFLO$_y$ state in the presence of transverse Zeeman field, with the parameters: $E_b/h=0.5$, $h_x/h=0.5$, $\alpha k_h/h=0.4$, $\mu/h=2.22$, $\Delta_Q/h\sim 0.32$, $Q_y/k_h\sim 0.86$. (d) Typical gapless contours in momentum space for an FFLO$_x$ state in the presence of transverse Zeeman field, with the parameters: $E_b/h=0.2$, $h_x/h=0.2$, $\alpha k_h/h=0.9$, $\mu/h=2$, $\Delta_Q/h\sim 0.19$, $Q_x/k_h\sim -0.11$.}
\label{fig4}
\end{figure}

With this understanding, we calculate the typical phase diagram for finite $h_x$ (see Fig. \ref{fig3}(a)). Apparently, the most striking effect of the transverse field is the stabilization of FFLO$_x$ states over a large parameter region, with the center-of-mass momentum of the FFLO$_x$ states opposite to the direction of the NIST SOC ($Q_x<0$). This is qualitatively consistent with the findings in Ref. \cite{chuanweifflo}, where a 3D Fermi gas with Rashba SOC and transverse fields was considered. When the SOC strength increases, the magnitude of $Q_x$ decreases exponentially as the Fermi surface asymmetry decreases, and the energy of the FFLO$_x$ state becomes exponentially close to that of an SF state (see Fig. \ref{fig4}(a)(b)). Similar scaling relations can be found with increasing $\mu$ \cite{seesupp}. As $\alpha$ or $\mu$ increases, when the magnitude of $Q_x$ becomes sufficiently small, the FFLO$_x$ state undergoes a continuous transition and becomes fully gapped. In the weak coupling limit, this gapped FFLO state (gFFLO$_x$) can be understood
as a pairing state within one helicity band, where the Fermi surface deformation induced by transverse field is accommodated by a finite center-of-mass momentum (see Fig. 3(b)). We expect that such an exotic pairing state is characteristic of the co-existence of SOC and Fermi surface asymmetry, and is not limited to systems under NIST SOC. On the other hand, the existence of various first-order boundaries and end points in Fig. \ref{fig3}(a) suggests stabilization of FFLO states with different $Q_x$ and $\Delta_Q$, consistent with our previous analysis. These FFLO$_x$ states eventually merge into a single FFLO$_x$ state beyond the end points. For finite $h_x$, we can still find considerable stability region for the FFLO$_y$ state, which is bounded from the FFLO$_x$ or gFFLO$_x$ states by first-order boundaries. The typical dispersion of the FFLO$_y$ state here has the symmetry $E^{\eta}_{k_x,k_y,\nu}=-E^{\eta'}_{-k_x,Q_y-k_y,\nu'}$. This is reflected in the contours of the gapless points (see Fig. \ref{fig4}(c)), which is qualitatively different from that of an FFLO$_x$ state (see Fig. \ref{fig4}(d)).

\emph{Conclusion}.--
We have studied the exotic phases in a 2D Fermi gas under the NIST SOC. The competition between the effective Zeeman fields and the anisotropic SOC leads to complicated phase structures, implying multiple phase separated states for a uniform gas and rich spatial phase structure for a trapped gas. Importantly, we demonstrate that various FFLO states, including a fully gapped FFLO state, can be stabilized under the NIST SOC. While the unique dispersion spectra and gapless contours of the FFLO states may serve as signatures for experimental detection, the various first-order phase transitions should also leave signatures in the {\it in-situ} density profiles of a trapped gas.

\emph{Acknowledgements}.--
We thank Lin Dong, Han Pu, and Vijay B. Shenoy for helpful discussions. This work is supported by NFRP (2011CB921200, 2011CBA00200), NNSF (60921091), NKBRP (2013CB922000), NSFC (11105134, 11274009), the Fundamental Research Funds for the Central Universities (WK2470000001, WK2470000006), and the Research Funds of Renmin University of China (10XNL016). W.Z. would also like to thank the NCET program for support.

\newpage
\begin{widetext}
\begin{appendix}
\section{Supplementary material}
In this supplementary material, we provide more details regarding the qualitative understanding and the properties of the FFLO$_x$ state under the NIST scheme.

As we have discussed in the main text, the various FFLO$_x$ states can be qualitatively understood as shifts of the local minima to finite $Q_x$ induced by Fermi surface asymmetries along the $x$-axis. This can be more easily understood by comparing the $Q=0$ phase diagram (see Fig. \ref{figsupp1}(a)) with the phase diagram in the main text (Fig. 3(a) therein). In the presence of a transverse Zeeman field, up to three local minima can be found in the thermodynamic potential for $Q=0$, corresponding to the normal, the nSF2 and the SF states, respectively (see Fig. \ref{figsupp1}(b)). This is reflected in the various first-order boundaries in Fig. \ref{figsupp1}(a). When we consider finite center-of-mass momentum pairing, these local minima will be shifted into the plane of finite $Q_x$. As a result, the first-order boundaries in Fig. 3(a) of the main text are reminiscent of the first-order boundaries for $Q=0$ in Fig. \ref{figsupp1}(a), but are shifted slightly. Of course, the phase diagram would be further modified by the competition from the FFLO$_y$ states, and the various local minima along $\mathbf{Q}=Q_x \hat{x}$ may merge as the parameters are tuned, which generates the end points we see in Fig. 3(a) of the main text. As an example, we plot in Fig. \ref{figsupp1}(c) a typical thermodynamic potential with $Q=0$ and in Fig. \ref{figsupp1}(d) the corresponding thermodynamic potential landscape when finite $Q_x$ pairing is considered. It appears that the original two local minima are both shifted into the finite $Q_x$ plane.

\begin{figure}
\includegraphics[width=6cm]{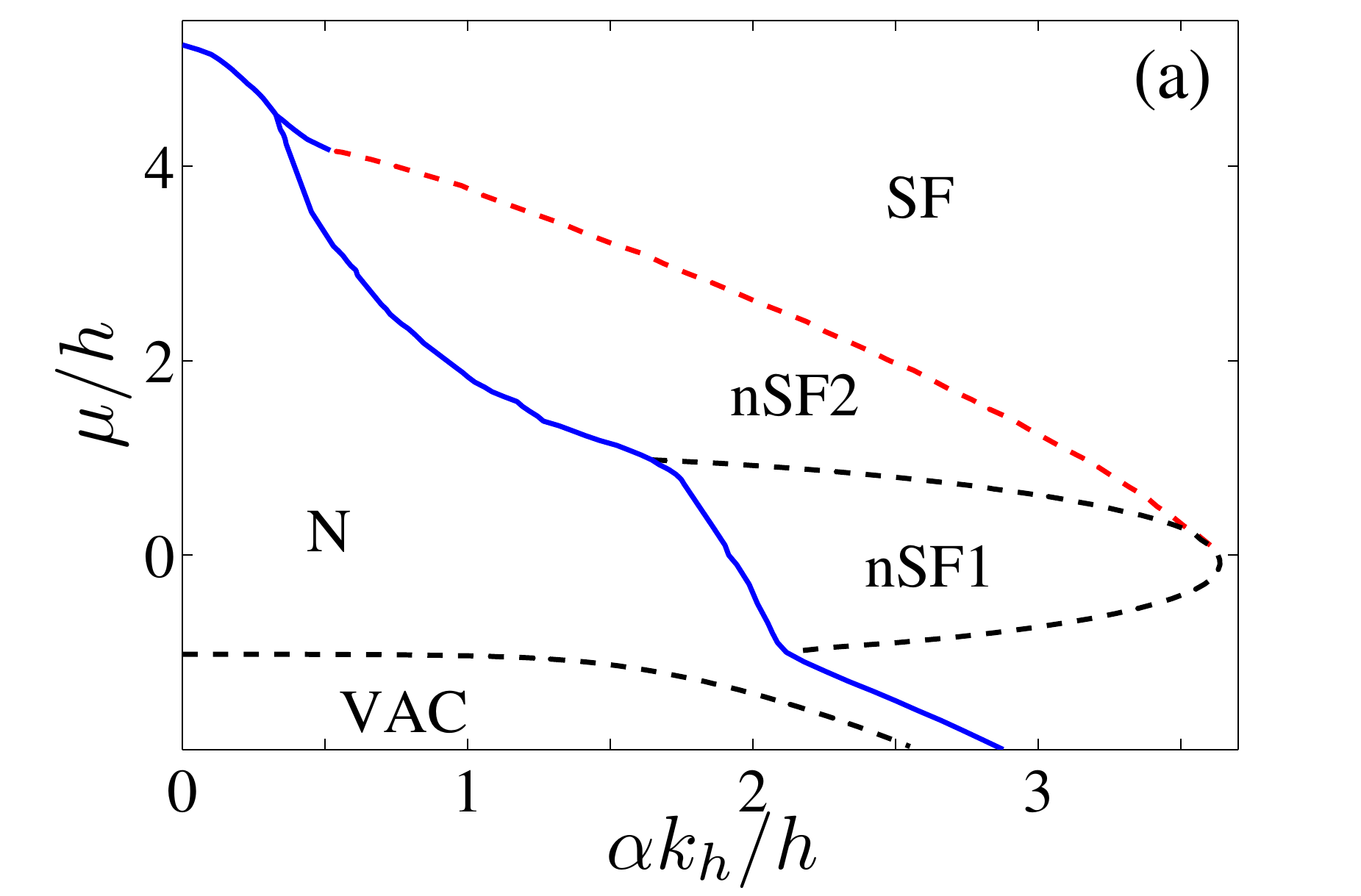}
\includegraphics[width=6cm]{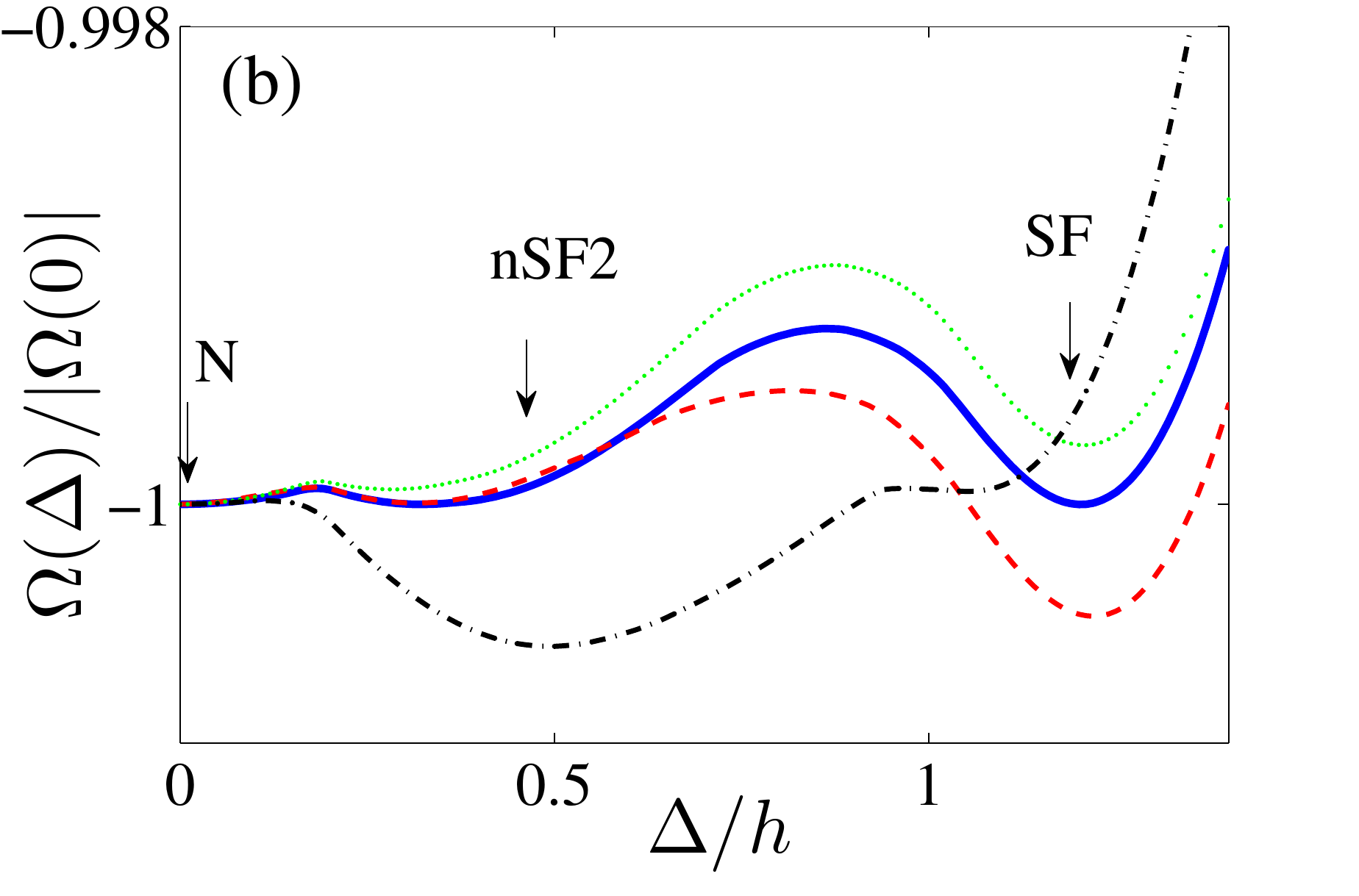}
\includegraphics[width=6cm]{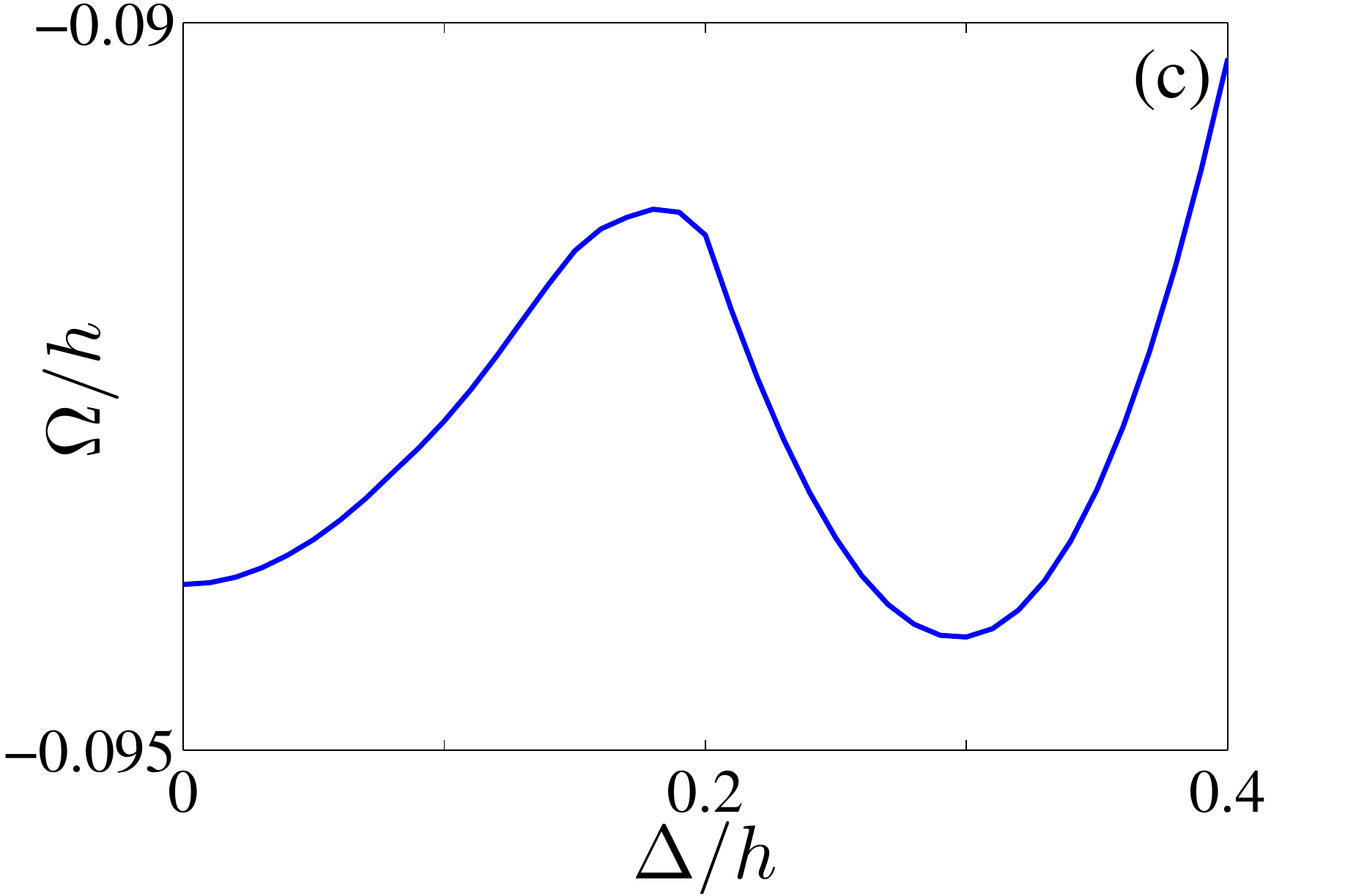}
\includegraphics[width=6cm]{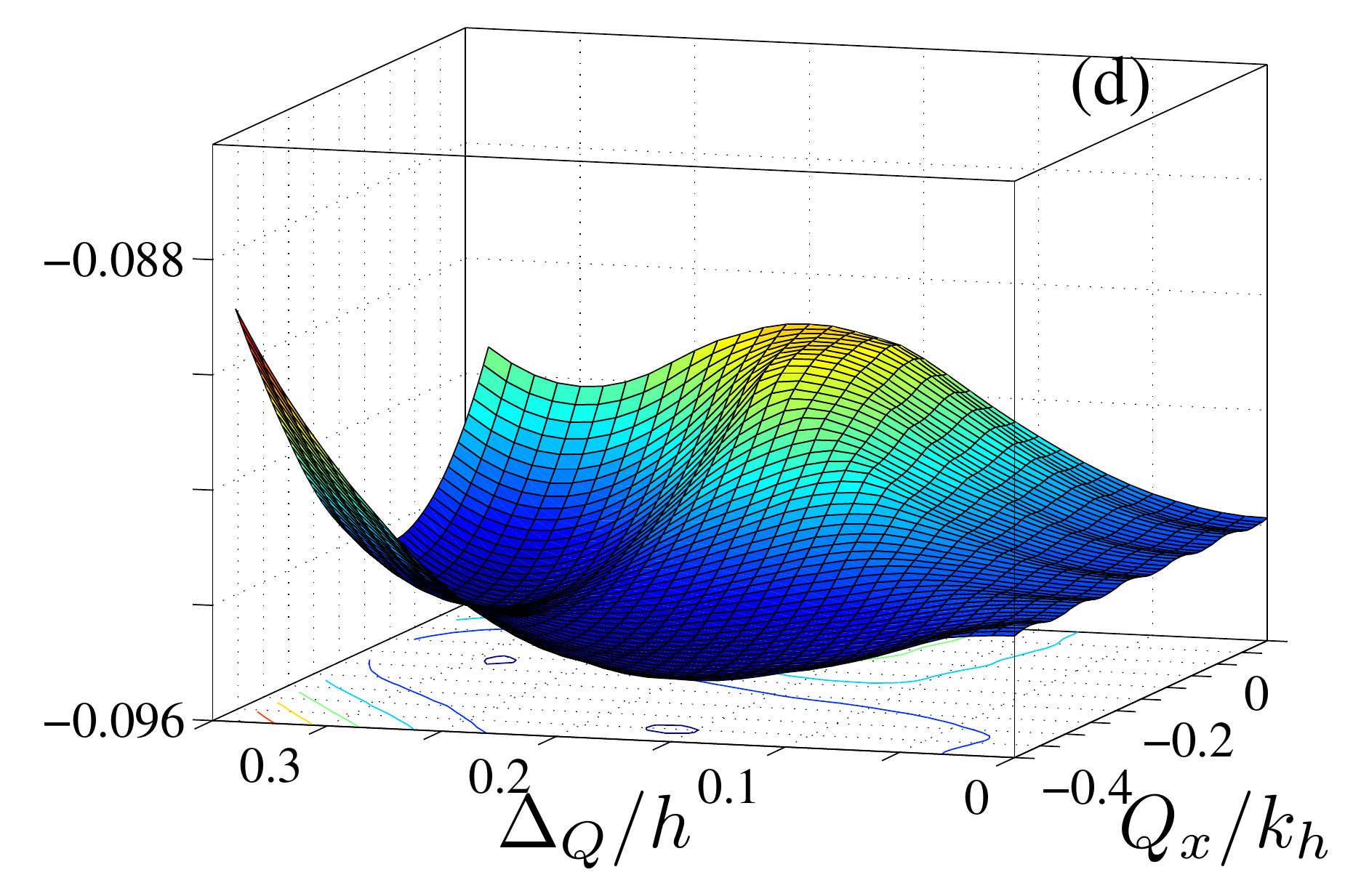}
\caption{(Color online) (a) Phase diagram on the $\mu$--$\alpha$ plane for $E_b/h=0.2$, $h_x/h=0.2$. Only zero center-of-mass momentum pairing states are considered here. The solid curves are first-order boundaries, while the dashed curves represent phase boundaries of continuous phase transitions.
(b) Typical thermodynamic potentials with up to three local minima, with the parameters:
$\alpha k_h/h=0.33$, $\mu/h=4.53$ (solid curve); $\alpha k_h/h=0.32$, $\mu/h=4.53$ (dotted curve); $\alpha k_h/h=0.42$, $\mu/h=4.2$ (dash-dotted curve); $\alpha k_h/h=0.33$, $\mu/h=4.6$ (dashed curve); while $E_b/h=0.2$, $h_x/h=0.2$ for all cases.
(c) Thermodynamic potential for $Q=0$ and $E_b/h=0.2$, $h_x/h=0.2$, $\alpha k_h/h\sim2.12$, $\mu/h\sim-0.98$. The local minima are located at $\Delta/h\sim 0.3$ and $\Delta/h=0$. (d) Thermodynamic potential as a function of $\Delta_Q$ and $Q_x$ for $E_b/h=0.2$, $h_x/h=0.2$, $\alpha k_h/h\sim2.12$, $\mu/h\sim-0.98$. The local minima are located at ($\Delta_{Q}/h\sim 0.29$, $Q_x/k_h\sim -0.11$) and ($\Delta_{Q}/h\sim 0.16$, $Q_x/k_h\sim -0.35$).}
\label{figsupp1}
\end{figure}

A remarkable feature of the phase diagram in the presence of a transverse Zeeman field and SOC is the enhanced stability region of the FFLO$_x$ state which seems to replace the SF state in the $Q=0$ phase diagram. As we have discussed in the main text, this is due to the Fermi surface deformation. As the Fermi surface deformation becomes smaller with increasing chemical potential or with increasing SOC strength, it is natural to expect that the energy gain by adopting a finite center-of-mass momentum pairing in the FFLO$_x$ state should become less. This can be seen by a direct scaling of the pairing parameters $Q_x$ and $\Delta_Q$ of the FFLO$_x$ state with increasing SOC strength (see Fig. 4(a) in the main text) or with increasing chemical potential (see Fig. \ref{figsupp2}(a)). Apparently, while the pairing order parameter $\Delta_Q$ keeps increasing with chemical potential and SOC strength, the magnitude of the center-of-mass momentum of the FFLO$_x$ state becomes exponentially small in the large $\mu$ and/or large $\alpha$ limit. Furthermore, the difference in thermodynamic potential between the SF state and the FFLO$_x$ state also becomes exponentially small. Therefore in real experiments, the parameter region where the FFLO$_x$ states can be observed should feature moderate SOC and moderate chemical potential.

\begin{figure}
\includegraphics[width=6cm]{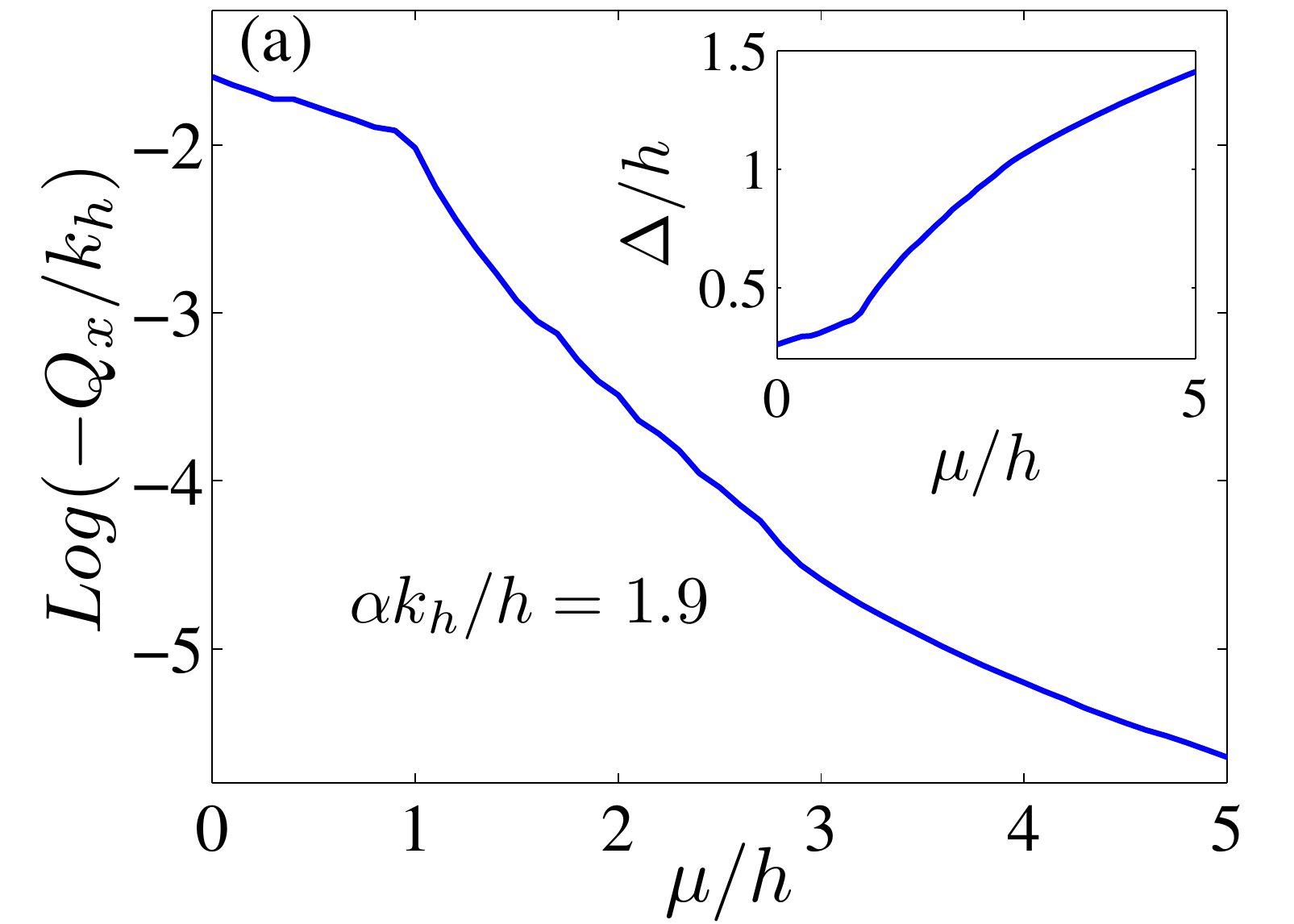}
\includegraphics[width=6cm]{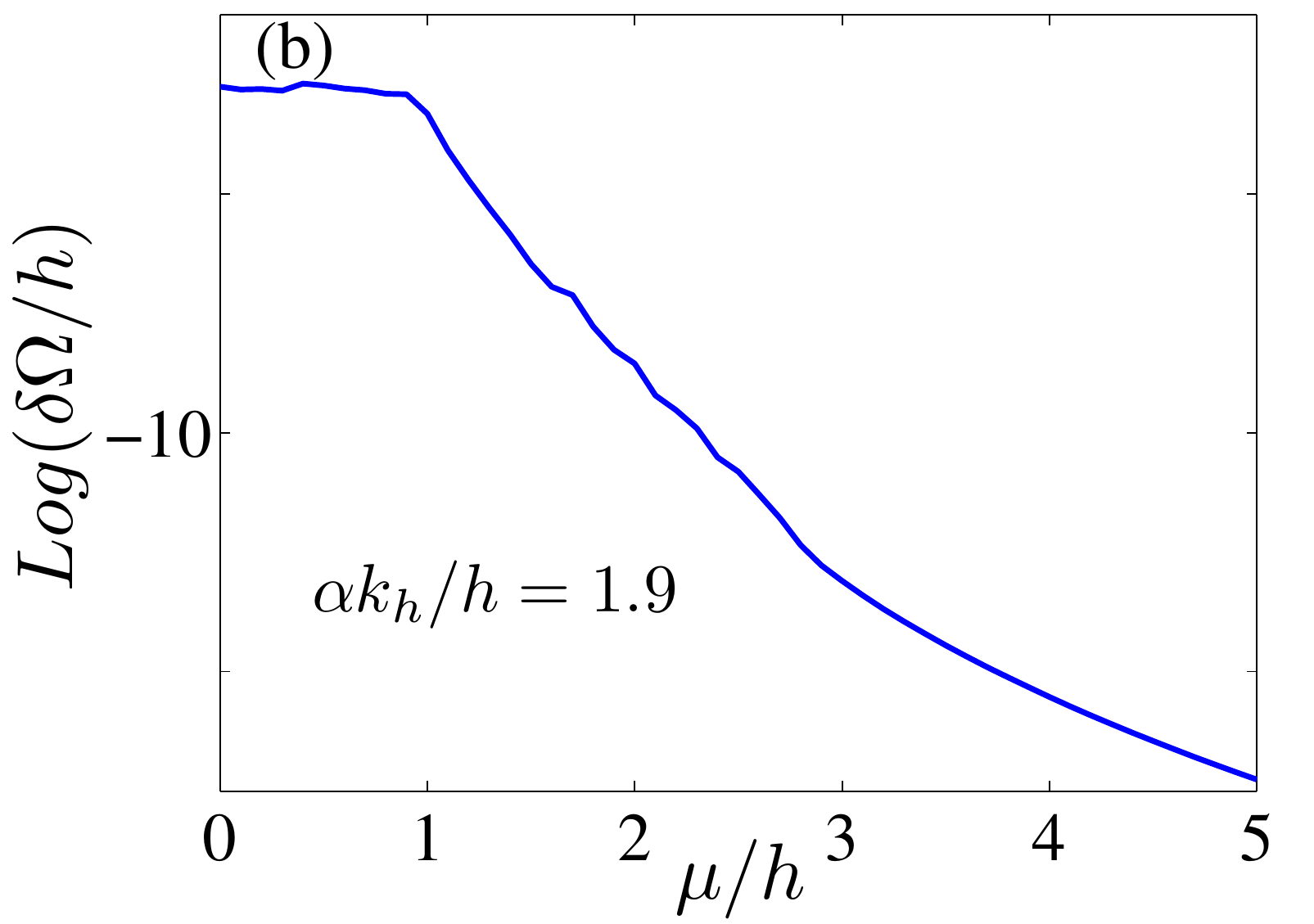}
\caption{(Color online) (a) Evolution of the center-of-mass momentum and the pairing order parameter (inset) for the ground state as the chemical potential increases. (b) Evolution of the difference in thermodynamic potential between the SF state and the FFLO$_x$ state with increasing chemical potential, with $\delta\Omega=\Omega_{\rm SF}-\Omega_{\rm FFLO_x}$. The parameters are: $E_b/h=0.2$, $h_x/h=0.2$, and $\alpha k_h/h=1.9$. }
\label{figsupp2}
\end{figure}

\end{appendix}
\end{widetext}


\begin{thebibliography}{99}
\bibitem{footnoteongaplesspoint} For the case of zero center-of-mass pairing and with no transverse field, it can be shown directly from the analytic form of the dispersion spectra that the gapless points can only occur on the $k_x=0$ axis.
\end{thebibliography}
\end{document}